# Effect of B-site bismuth doping on magnetic and transport properties of La$_{0.5}$Ca$_{0.5}$Mn$_{1-x}$Bi$_x$O$_3$ thin films

Himanshu Sharma[1,2*], Deepak Kumar[1], Ashwin Tulapurkar[3], C. V. Tomy[1]

[1]Department of Physics, Indian Institute of Technology Bombay, Powai, Mumbai – 400076, India.
[2]Institute for Materials Research, Tohoku University, Sendai 980-8577, Japan.
[3]Department of Electrical Engineering, Indian Institute of Technology Bombay, Powai, Mumbai – 400076, India.
E-mail: * himsharma@imr.tohoku.ac.jp

## Abstract

The magnetic properties in manganite have been found to be highly sensitive to the doping and structural manipulations. Here, we report the effect of B-site bismuth doping on the magnetic and transport properties in La$_{0.5}$Ca$_{0.5}$Mn$_{1-x}$Bi$_x$O$_3$ (LCMBO) thin films (for x = 0, 0.02 and 0.05) for high-efficiency spintronics devices. For thin film of LCMBO (with x = 0.02), a significant increase in the magnetization and ferromagnetic ordering temperature ($T_C$) are observed. Also, about 98% magnetoresistance (MR) and unusually large (~ 42%) anisotropic magnetoresistance (AMR) is observed at 50 K in the same LCMBO (for x = 0.02) thin film. This observed improvement in $T_C$, MR and AMR in LCMBO (with x = 0.02) thin film may be attributed to the modulation of the trapped electrons through JT-distortions due to the replacement of Mn$^{+3}$ ions by larger Bi$^{+3}$ ions. With further increase in bismuth doping (for x = 0.05) at the B-site, a significant decrease in magnetization and $T_C$ have been observed.

## Introduction

Mixed valent doped perovskite manganites of the type $R_{1-x}A_x$MnO$_3$ ($R$ = trivalent rare earth metal - La, Pr or Nd, etc; $A$ = divalent alkali earth metal - Ca, Ba or Sr, etc) exhibit a variety of physical phenomena (e.g., colossal magnetoresistance (CMR), magnetocaloric effects (MCE), charge/orbital ordering, complex magnetic ordering, etc) [1-4]. These physical phenomena are attributed due to the simultaneous interaction of several physical parameters, viz., spin, charge, orbital and lattice [1-9]. Also, these most interesting properties are revealed by the complex mixed valent doped perovskites due to the substitution of trivalent (+3) cation of a rare earth ($R^{+3}$) with a divalent alkali earth cation ($A^{+2}$), which results a charge imbalance in the system [1-9]. This charge imbalance is compensated by the Mn cations by transferring part of the manganese cation valence state from +3 to +4 [1-9]. This deviation in the 3d valence state of transition metal (Mn) derives the Jahn-Teller (J-T) distortions [1,2,7] and double exchange (DE) interaction [1,2,7] between the Mn-ions. Also, it is reported in ferromagnetic (FM) and charge ordered antiferromagnetic (CO-AFM) phase coexisting $La_{0.5}Ca_{0.5}MnO_3$ compound that the substitution of diamagnetic element like bismuth (Bi) at B-site also induces important changes in its physical properties [5-6]. However, the effect of Bi-doping at B-site is still not investigated in its thin films, which will be fascinating due to its possible applications in spintronics. Further, the replacement of $Mn^{+3}$ ions by larger $Bi^{+3}$ will affect the Spin-orbit coupling associated with DE-interaction, which can gives augment to the magnetic anisotropy in $La_{0.5}Ca_{0.5}Mn_{1-x}Bi_xO_3$ (LCMBO) thin films [10-16]. Thus the modulation of magnetic properties, transport properties and anisotropic magnetoresistance (AMR) of these materials can be effectively realized by Bi-doping at B-site. Also, remarkable research efforts have been dedicated to improve the transport properties (i.e., to obtain large values of AMR and MR) in order to improve the sensitivity of its use devices like magnetic field sensors, low-cost inertial navigation systems, magnetoresistive random access memory (MRAM) and switching devices [10-19].

Here, we present a systematic study to investigate the effect of B-site bismuth doping on the magnetic properties, transport properties and magnetic anisotropy (i.e., anisotropic magnetoresistance) in $La_{0.5}Ca_{0.5}Mn_{1-x}Bi_xO_3$ (LCMBO) thin films (for x = 0, 0.02 and 0.05) grown on $LaAlO_3$ (LAO) (001) substrate.

**Materials and methods**

Dense targets of $La_{0.5}Ca_{0.5}Mn_{1-x}Bi_xO_3$ (LCMBO) with x = 0, 0.02 and 0.05 were prepared in air by the standard solid-state reaction method using powders of constituent oxides, $La_2O_3$, $MnO_2$, $Bi_2O_3$ and $CaCO_3$ of purity > 99.9% as the starting materials. The materials were mixed in proper stoichiometry and grounded well before giving the heat treatment at 1200ºC for 36 hours in alumina crucibles. The mixture was ground well once again just after the heat treatment. The final sintering of the pelletized powder was done at 1350ºC for 48 hours. These densified pellets of $La_{0.5}Ca_{0.5}Mn_{1-x}Bi_xO_3$ (LCMBO), for x = 0, 0.02 and 0.05 ware used as target materials during thin film deposition process. From these sintered targets thin films of $La_{0.5}Ca_{0.5}Mn_{1-x}Bi_xO_3$ (LCMBO), for x = 0, 0.02 and 0.05 with same thicknesses (50 nm) were deposited on $LaAlO_3$ (LAO) (001) substrates, using a KrF-Pulsed Laser Deposition (PLD) system with $\lambda = 248$ nm and energy density of ~ 2 $Jcm^{-2}$ [10,16-20]. The oxygen pressure and substrate temperature through the deposition were $3.5 \times 10^{-1}$ mbar and 700°C, respectively [10]. After the deposition, PLD chamber was vented to atmospheric oxygen pressure (~1 bar) and then cooled down to room temperature at a rate of 10° C/min [10].

The thicknesses of the thin films were measured by the profilometer (DektakXT stylus surface profiler), which has the vertical resolution: ≤ 0.1 nm. The thin films were patterned in 100 μm × 2 mm rectangular channel using optical lithography and ion beam milling process. The gold pads were created for four probe transport (electrical resistivity) measurements using optical lithography followed by "Au" deposition (by sputtering process) and lift-off.

The magnetization measurements were performed using a SQUID Magnetometer (SQUID-VSM, Quantum Design Inc., USA). The magnetization was recorded as a function of temperature and magnetic field. The physical property measurement system (PPMS - Quantum Design Inc., USA) was used for transport measurements as a function of temperature and magnetic field. Also, the angular dependent magnetoresistance or anisotropic magnetoresistance (AMR) measurements as a function of temperature and magnetic field were performed using a PPMS system and a sample rotator assembly (motor controlled Horizontal rotator-Quantum Design Inc.), in order to rotate the sample continuously during the measurement.

## Results and discussions

### X-ray diffraction

Figure 1 shows the XRD pattern of the $La_{0.5}Ca_{0.5}Mn_{1-x}Bi_xO_3$ (LCMBO) thin films for x = 0, 0.02 and 0.05. A strong (001) and (002) peak (marked as "*") of LCMBO could be clearly recognized from the XRD data, which confirms the single phase nature of the thin films. Further, from the XRD data only the (00$l$) reflections of the LCMBO thin films are present along with the reflections from the LaAlO$_3$ (LAO) substrate, indicating c-axis growth. Further, inset of Fig. 1 shows a detailed spectrum near the (002) peak of LAO-substrate and the thin films, which highlight the significant shifts of (00$l$) reflections of the LCMBO thin films.

Also, the d-spacing of the samples can be calculated using the Bragg's equation of XRD [1,17]:

$$\lambda = \frac{2d}{\sqrt{h^2 + k^2 + l^2}} \sin\theta \quad ----- (1)$$

where, λ is the wavelength of the X-ray source (Cu-Kα, λ = 1.54 Å), "2θ" is the Bragg angle, d is the lattice parameter and ($h,k,l$) are the miller indices.

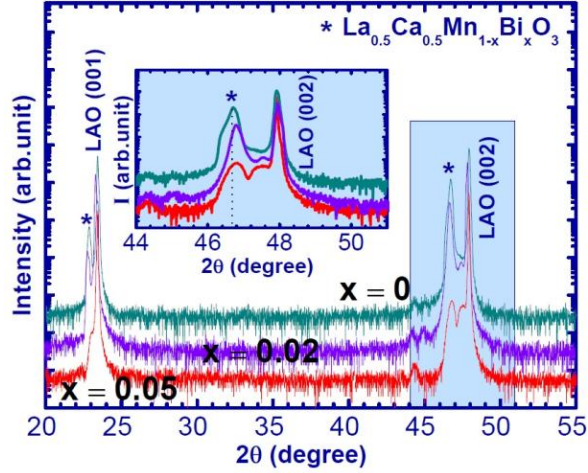

**Figure 1** X-ray diffraction patterns of the La$_{0.5}$Ca$_{0.5}$Mn$_{1-x}$Bi$_x$O$_3$ (LCMBO) thin films with x = 0, 0.02 and 0.05 grown on LAO (001) substrate. Inset shows a detailed spectrum in the region of the (002) reflections of LAO-substrate and the thin films.

From Eq. (1), the d-spacing for all the thin films along with LAO (001) substrate is calculated as shown in Table 1. The relative change in the d-spacing with respect to that of LAO substrate (3.791Å) is measured using (($d_{LAO}$-$d_{LCMBO}$)/$d_{LAO}$) to highlight the change in strain on the film surface with Bi-doping (see Table 1). Such a small variation in d-spacing can result in large change in the strain on the film surface. Here the variation in d-spacing is possibly associated with the increase in unit cell volume due to replacement of Mn$^{+3}$ ions by larger Bi$^{+3}$.

**Table 1**
The calculated d-spacing and the relative change in the d-spacing of La$_{0.5}$Ca$_{0.5}$Mn$_{1-x}$Bi$_x$O$_3$ thin films with respect to that of LAO substrate (3.791Å).

| | x | d-spacing (Å) | $d_R = (d_{LAO}$-$d_{LCMBO})/d_{LAO}$ |
|---|---|---|---|
| La$_{0.5}$Ca$_{0.5}$Mn$_{1-x}$Bi$_x$O$_3$ | 0 | 3.88708 | − 0.02534 |
| | 0.02 | 3.87978 | − 0.02341 |
| | 0.05 | 3.88464 | − 0.02470 |

## Magnetization

The temperature dependence of the field-cooled (FC) magnetization of the La$_{0.5}$Ca$_{0.5}$Mn$_{1-x}$Bi$_x$O$_3$ (LCMBO) thin film for x = 0, 0.02 and 0.05 in an applied magnetic field of 100 Oe is shown in Fig. 2(a). For a detailed study, one curve of zero field-cooled (ZFC) magnetization of the $La_{0.5}Ca_{0.5}Mn_{1-x}Bi_xO_3$ (LCMBO) thin films for x = 0.02 in an applied magnetic field of 100 Oe is also shown in Fig. 2(a). The derivatives of FC-magnetization to highlight the shift in the ferromagnetic Curie temperature ($T_C$) are shown in Fig. 2(b). Inset of Fig. 2(b) shows the variations of $T_C$ (the peak temperatures in the derivative plot) as a function of bismuth doping (x).

It is to be noted that a large increase in magnetization with a significant shift in ferromagnetic transition temperature ($T_C$) is observed in $La_{0.5}Ca_{0.5}Mn_{0.98}Bi_{0.02}O_3$ thin film in comparison with that of $La_{0.5}Ca_{0.5}MnO_3$ thin film. Whereas, the further decrease of manganese amount in the thin film (i.e., $La_{0.5}Ca_{0.5}Mn_{0.95}Bi_{0.05}O_3$ with 5% Bi-doping on B-site) results a large decrease in magnetization and $T_C$ than those recorded for the $La_{0.5}Ca_{0.5}MnO_3$ thin film. The magnetization results observed (i.e., increase/decrease in magnetization and shift in $T_C$) for the thin films of $La_{0.5}Ca_{0.5}Mn_{1-x}Bi_xO_3$ (LCMBO) with x = 0, 0.02 and 0.05 are consistent with the results reported by A. Krichene et al. [5-6], for their respective bulk samples. The ferromagnetic nature of the $La_{0.5}Ca_{0.5}Mn_{1-x}Bi_xO_3$ (LCMBO) thin films is further confirmed through the magnetization measurements as a function of magnetic field at 5 K, which is shown in Fig. 3.

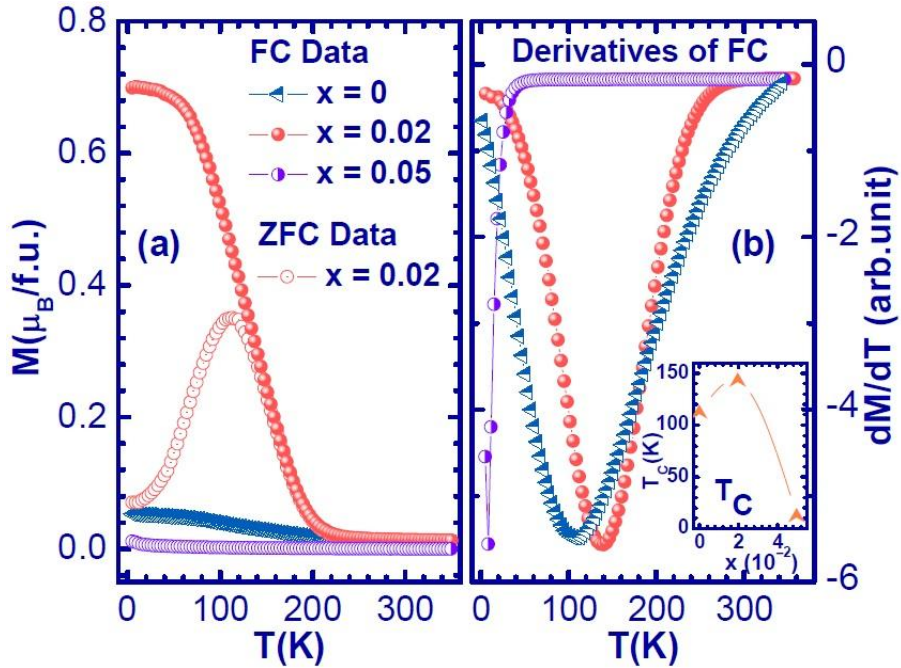

**Figure 2** (a) Temperature dependence of FC magnetization in an applied magnetic field (H) of 100 Oe for $La_{0.5}Ca_{0.5}Mn_{1-x}Bi_xO_3$ (for x = 0, x = 0.02 and x = 0.05) thin films and ZFC magnetization of $La_{0.5}Ca_{0.5}Mn_{0.98}Bi_{0.02}O_3$ thin film measured with H parallel to the film plane, (b) The derivatives of FC magnetization to highlight the shift in $T_C$. Inset shows the variation of $T_C$ as a function of x.

It is evident that for La$_{0.5}$Ca$_{0.5}$Mn$_{1-x}$Bi$_x$O$_3$ (LCMBO) thin films, we see a saturation magnetization, confirming the ferromagnetic nature of these (For x = 0, 0.02 and 0.05) films at 5 K (see Fig. 3 and its inset for x = 0.05).

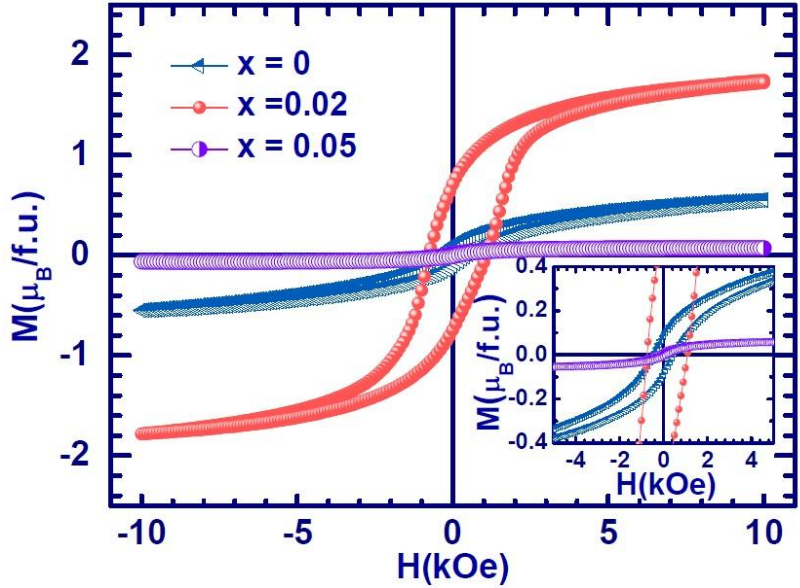

**Figure 3** Magnetization (M) curves as a function of magnetic field (H) for La$_{0.5}$Ca$_{0.5}$Mn$_{1-x}$Bi$_x$O$_3$ (for x = 0, x = 0.02 and x = 0.05) thin films at 5K, measured with H parallel to the film plane.

## Magnetoresistance

Figures 4(a) and 4(b) show the variation of resistivity as a function of temperature with different applied magnetic field for thin films of La$_{0.5}$Ca$_{0.5}$MnO$_3$ and La$_{0.5}$Ca$_{0.5}$Mn$_{0.98}$Bi$_{0.02}$O$_3$, respectively. Also, inset in Fig. 4(b) shows the resistivity data of La$_{0.5}$Ca$_{0.5}$Mn$_{0.95}$Bi$_{0.05}$O$_3$ thin film. It is observed from the transport data that thin film of La$_{0.5}$Ca$_{0.5}$MnO$_3$ shows the insulating behaviour even in the applied magnetic field of 9 T. On the other hand La$_{0.5}$Ca$_{0.5}$Mn$_{0.95}$Bi$_{0.05}$O$_3$ (i.e., 2% Bi-doping on B-site) thin film shows a large decrease in resistivity with applied magnetic field below ferromagnetic ordering temperature.

In addition to highlight the semiconducting behavior of La$_{0.5}$Ca$_{0.5}$Mn$_{1-x}$Bi$_x$O$_3$ (for x = 0, x = 0.02 and x = 0.05) thin films at higher temperature, the Emin–Holstein theory of the adiabatic small polaron hopping (ASPH) model [18] is used, which is expressed as:

$$\rho(T) = BT \exp(E_a/k_B T) \quad \text{-----------} \quad (2)$$

where $E_a$ is the activation energy for hopping conduction electrons and B is the residual resistivity. Figure 4d shows the results observed for the ASPH model and the straight line (in colors) is a fit to Eq. (2).

Figure 5 (a) shows the MR plots for $La_{0.5}Ca_{0.5}Mn_{1-x}Bi_xO_3$ (for x = 0, x = 0.02 and x = 0.05) thin films. The percentage MR is calculated using the formula, $([\rho(0) - \rho(H)]/\rho(0)) \times 100$; where $\rho(0)$ and $\rho(H)$ (here, H = 50 kOe ) are the resistivity in zero magnetic field and applied magnetic field of H, respectively. The large MR is observed in the $La_{0.5}Ca_{0.5}Mn_{0.98}Bi_{0.02}O_3$ thin film in comparison with the MR values in $La_{0.5}Ca_{0.5}MnO_3$ thin film. More than 95% MR is observed below ferromagnetic transition temperature ($T_C$) in $La_{0.5}Ca_{0.5}Mn_{0.98}Bi_{0.02}O_3$ thin film. The observed decrease in resistivity with applied magnetic field is associated with the double exchange (DE) interactions [1,5], which is further confirmed through magnetic field dependent MR curve of $La_{0.5}Ca_{0.5}Mn_{0.98}Bi_{0.02}O_3$ thin film at 30 K (see Fig. 5 (b)), where we not only observed a magnetic coercivity but also a saturation behavior. This confirms the ferromagnetic ordering of conduction electrons ($e_g$) in magnetic domains, which favors the DE-interactions.

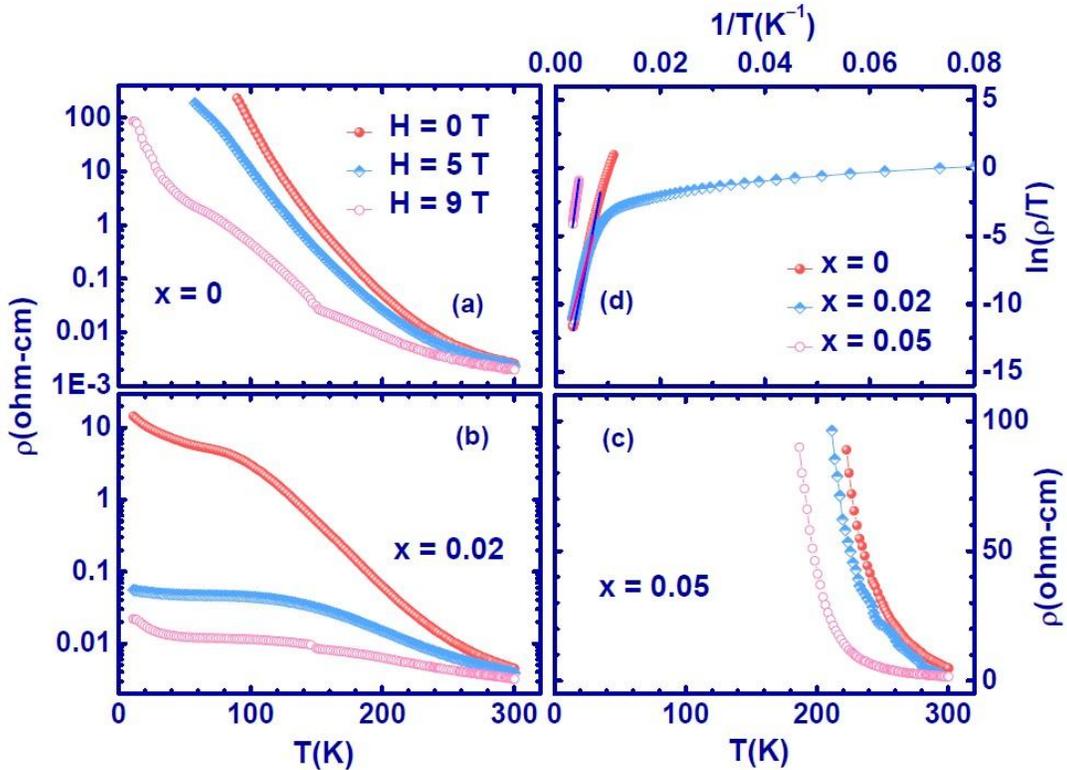

**Figure 4** Temperature dependence of the resistivity for $La_{0.5}Ca_{0.5}Mn_{1-x}Bi_xO_3$ /LAO thin films for (a) x = 0, (b) x = 0.02 and (c) x = 0.05 in different applied magnetic field, (d) ln($\rho$/T ) Vs 1/T plots for x = 0, 0.02 and 0.05 to highlight the semiconductor behavior at higher temperatures. Line is the fit to Eq. (2).

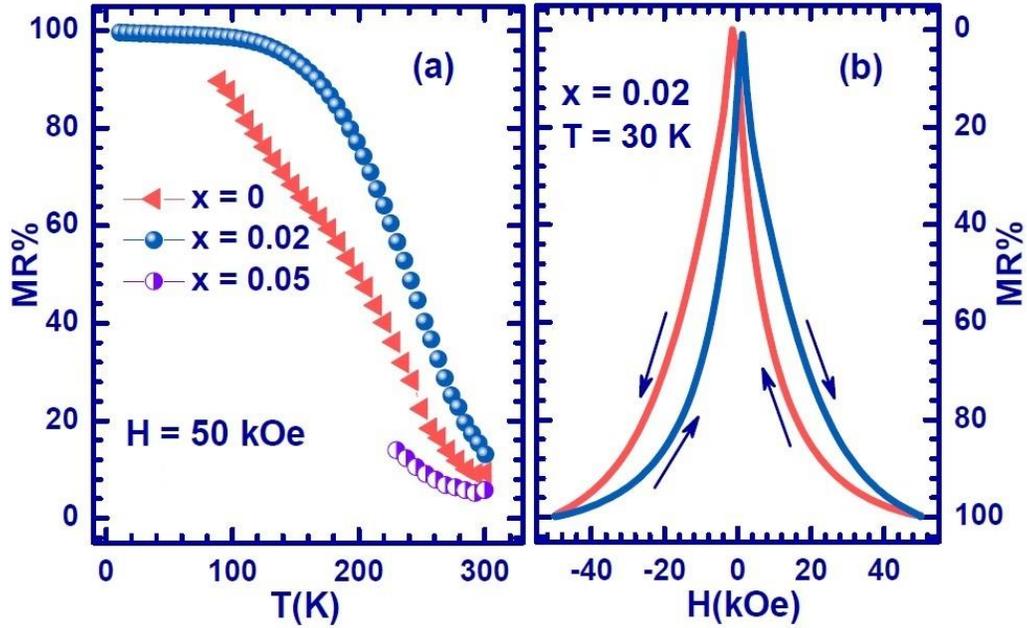

**Figure 5** (a) Temperature dependence of MR for La$_{0.5}$Ca$_{0.5}$Mn$_{1-x}$Bi$_x$O$_3$ thin films, (for x = 0, 0.02 and 0.05) for applied magnetic field of 50 kOe, (b) MR curve of La$_{0.5}$Ca$_{0.5}$Mn$_{0.98}$Bi$_{0.02}$O$_3$ thin film at 30 K.

The transport results (see Fig. 4 and 5) in consistent with the magnetization measurements (see Fig. 2 and 3) confirms that 2% Bi-doping on B-site in La$_{0.5}$Ca$_{0.5}$Mn$_{1-x}$Bi$_x$O$_3$ (LCMBO) is sufficient to enhance ferromagnetism (FM) and hence the conductivity. A limited amount of Bi-doping results a decrease in Mn$^{+3}$ content that can induce a FM through the enhancement of DE-interactions [1,5-6].

**Anisotropic Magnetoresistance**

Anisotropic magnetoresistance (AMR) of the La$_{0.5}$Ca$_{0.5}$MnO$_3$ and La$_{0.5}$Ca$_{0.5}$Mn$_{0.98}$Bi$_{0.02}$O$_3$ thin films were measured by varying the angle (θ) from 0$^o$ to 360$^o$ at different temperatures for an applied magnetic field of 50 kOe using a four probe method, where θ = 0$^o$ corresponds to the configuration of H perpendicular to the plane of the film and 90$^o$ corresponds to H parallel to the plane of the film.

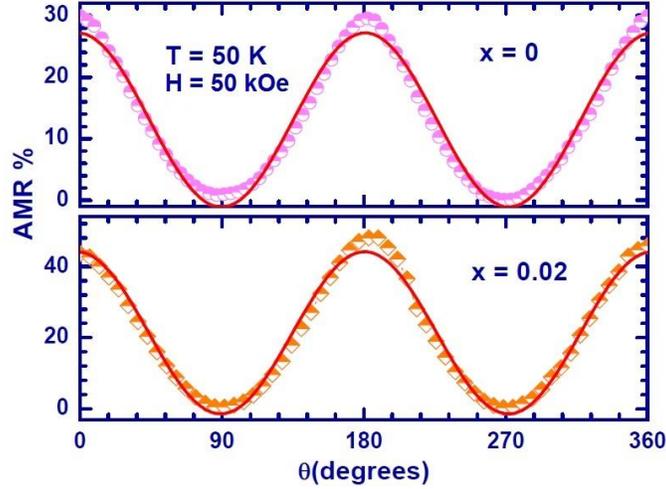

**Figure 6** Anisotropic Magnetoresistance (AMR) for La$_{0.5}$Ca$_{0.5}$Mn$_{1-x}$Bi$_x$O$_3$ thin films for x = 0 and 0.02 at 50 K with an applied magnetic field of 50 kOe. The line shows the fit to cos$^2$ θ.

Figure 6, shows the percentage AMR as a function of sample position (θ) at 50 K for La$_{0.5}$Ca$_{0.5}$Mn$_{1-x}$Bi$_x$O$_3$ thin films (with x = 0 and 0.02) with an applied magnetic field of 50 kOe. Where, the percentage AMR is defined as [{ρ(θ°) − ρ(90°)}/ρ (90°)]×100 [10]. The observed AMR for La$_{0.5}$Ca$_{0.5}$MnO$_3$ and La$_{0.5}$Ca$_{0.5}$Mn$_{0.98}$Bi$_{0.02}$O$_3$ thin film shows two-fold cos$^2$θ dependence symmetry at 50 K (see Fig. 6). The line (in red) shows the fit to cos$^2$ θ. Here, the observed maxima at θ = 0° and a minima at θ = 90° is due to the fact that the easy magnetization direction (easy axis) for both of the films is θ = 90° (parallel to the plane of the film).

Next, we have measured the maximum AMR [10] (defined as [{ρ(0°) − ρ(90°)}/ρ (90°)]×100) as a function of temperature for La$_{0.5}$Ca$_{0.5}$Mn$_{1-x}$Bi$_x$O$_3$ thin films (for x = 0, 0.02 and 0.05) thin films (see Fig. 7) with an applied magnetic field of 50 kOe. Large values of AMR have been observed in La$_{0.5}$Ca$_{0.5}$Mn$_{0.98}$Bi$_{0.02}$O$_3$ thin film in comparison with that of La$_{0.5}$Ca$_{0.5}$MnO$_3$ thin film. Also, AMR increases below ferromagnetic ordering temperature ($T_C$). It was also evident that 2% Bi-doping on B-site in La$_{0.5}$Ca$_{0.5}$Mn$_{1-x}$Bi$_x$O$_3$ (LCMBO) enhance the anisotropy of the thin film. Whereas, with further increase in Bi-doping on B-site in La$_{0.5}$Ca$_{0.5}$Mn$_{1-x}$Bi$_x$O$_3$ (LCMBO) (i.e., 5% Bi-doping) decrease the anisotropy of the thin film.

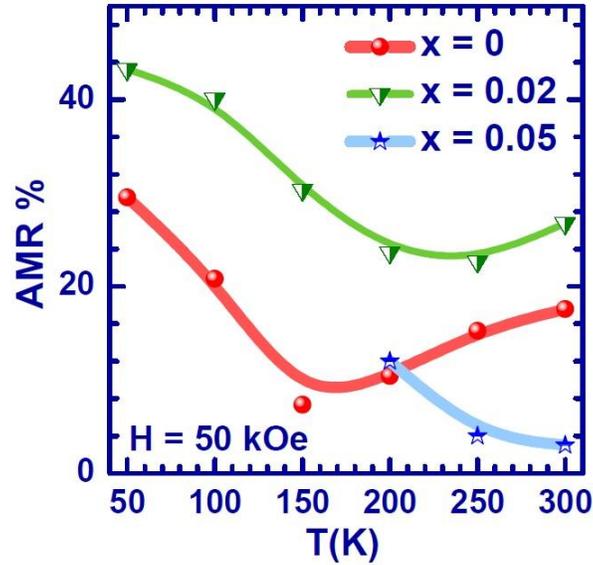

**Figure 7** Temperature dependence of AMR for La$_{0.5}$Ca$_{0.5}$Mn$_{1-x}$Bi$_x$O$_3$ thin films, (for x = 0, 0.02 and 0.05) with applied magnetic field of 50 kOe.

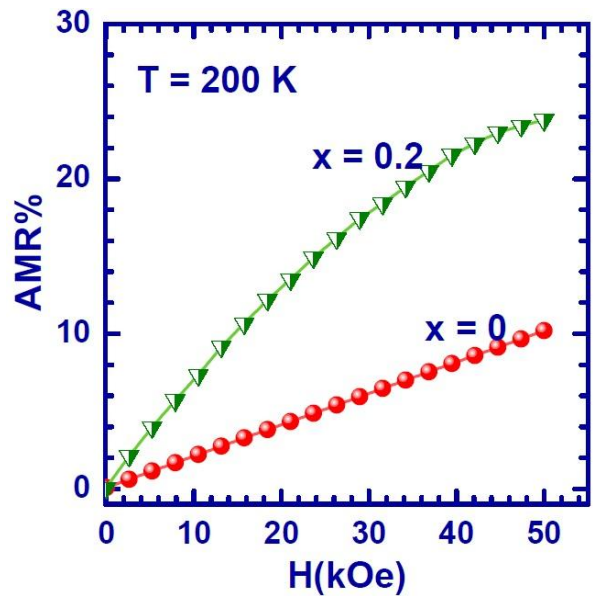

**Figure 8** Magnetic Field dependence of AMR for La$_{0.5}$Ca$_{0.5}$Mn$_{1-x}$Bi$_x$O$_3$ thin films, (for x = 0 and 0.02) at 200 K.

In order to investigate the magnetic field effect on AMR, we have measured the maximum AMR as a function of magnetic field at 200 K, for La$_{0.5}$Ca$_{0.5}$MnO$_3$ and La$_{0.5}$Ca$_{0.5}$Mn$_{0.98}$Bi$_{0.02}$O$_3$ thin films, which is shown in Fig. 8. The enhancement in anisotropy of La$_{0.5}$Ca$_{0.5}$Mn$_{0.98}$Bi$_{0.02}$O$_3$ thin film is further confirmed by magnetic field dependent AMR data as shown in Fig. 8.

It is shown in Fig. 7 that the maximum AMR for La$_{0.5}$Ca$_{0.5}$Mn$_{0.98}$Bi$_{0.02}$O$_3$ and La$_{0.5}$Ca$_{0.5}$MnO$_3$ thin films have been observed at 50 K. The maximum AMR observed for La$_{0.5}$Ca$_{0.5}$Mn$_{0.98}$Bi$_{0.02}$O$_3$ thin film at 50 K is 42%. However, maximum AMR of 30% is observed in La$_{0.5}$Ca$_{0.5}$MnO$_3$ thin film at 50 K.

Also, the observed two-fold cos$^2\theta$ dependence symmetry of AMR is consistent for all the temperatures and it is further confirmed through AMR measurement of La$_{0.5}$Ca$_{0.5}$Mn$_{0.98}$Bi$_{0.02}$O$_3$ thin film at different temperatures with an applied magnetic field of 5 T as shown in Fig. 9. Also, the angular dependence of magnetoresistance for La$_{0.5}$Ca$_{0.5}$Mn$_{0.98}$Bi$_{0.02}$O$_3$ thin film plotted in polar coordinates highlights the symmetry of the curve as shown in Fig. 9 b.

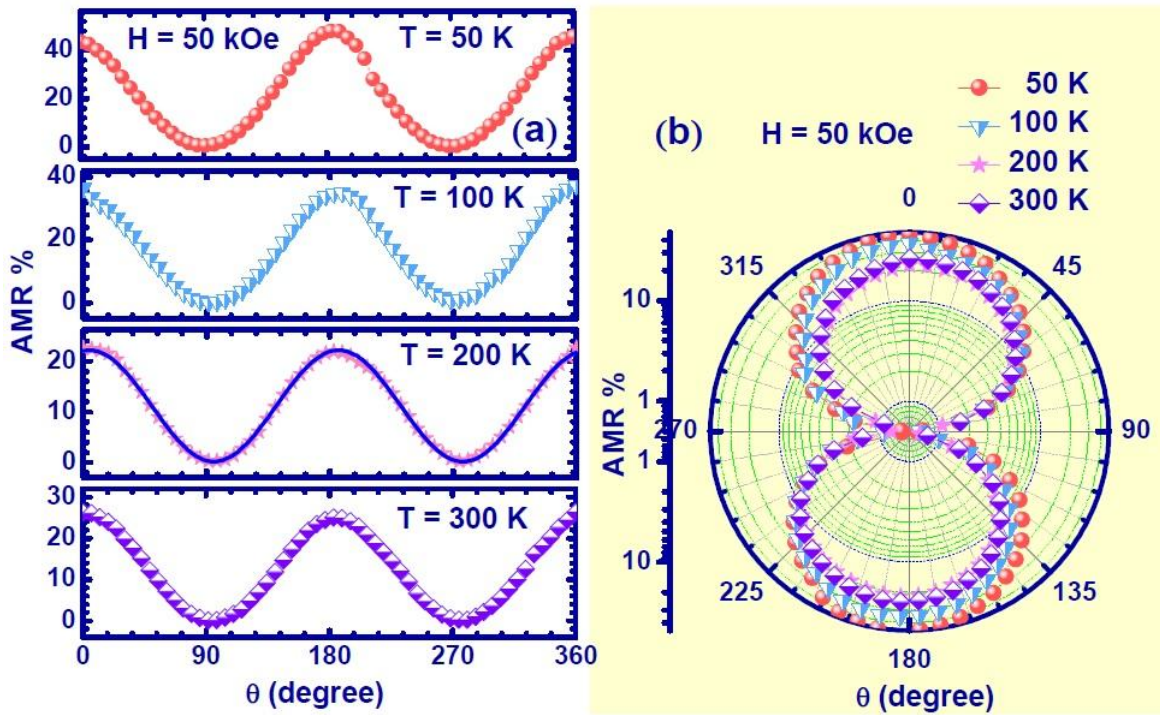

**Figure 9** Anisotropic Magnetoresistance (AMR) of La$_{0.5}$Ca$_{0.5}$Mn$_{0.98}$Bi$_{0.02}$O$_3$ thin film at different temperatures with an applied magnetic field of 5 T, (a) plotted in linear coordinates and the line shows the fit to cos$^2\theta$, and (b) plotted in polar coordinates.

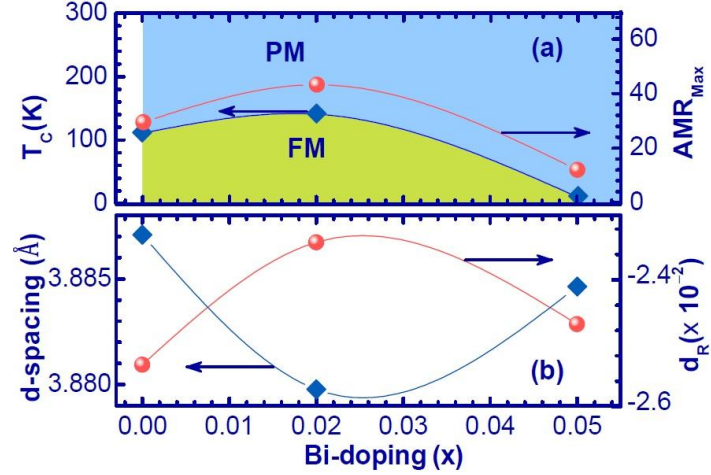

**Figure 10** (a) B-site bismuth doping (x) dependence of $T_C$ (left axis) and maximum AMR (right axis), (b) B-site bismuth doping (x) dependence of d-spacing (left axis) and relative shift in the d-spacing (right axis) with respect to that of LAO substrate (3.791Å) of $La_{0.5}Ca_{0.5}Mn_{1-x}Bi_xO_3$ thin films.

The results obtained in this paper in conjunction with the results from A. Krichene et al. [5-6] (for bulk samples) bring in interesting aspects regarding the enhancement of ferromagnetic ordering, magnetoresistance (MR) and anisotropic magnetoresistance (AMR) in $La_{0.5}Ca_{0.5}Mn_{1-x}Bi_xO_3$ thin films. In the mixed valent doped perovskite manganites electrons are trap due to Jahn-Teller (JT) distortions [1-9], whereas the Zener double exchange (DE) mechanism [1-9] try to hop the charge carriers ($e_g$ electrons) through $Mn^{+3}$ to $Mn^{+4}$ [1-9]. The interplay between the trapping of electrons due to JT-distortions and the hopping of charge carriers due to the DE-mechanism determined the transport and magnetic nature of the manganites [1-9]. Thus the observed improvement in magnetization, magnetoresistance (MR) and anisotropic magnetoresistance (AMR) in $La_{0.5}Ca_{0.5}Mn_{0.98}Bi_{0.02}O_3$ is possibly associated with the variation of strain due to change in unit cell volume through the replacement of $Mn^{+3}$ ions by larger $Bi^{+3}$. We have further surmised this in Fig. 10 (a) and (b), where the variation in d-spacing (to highlight the variation of strain in LCMBO thin film), ferromagnetic ordering temperature ($T_C$) and maximum anisotropic magnetoresistance (AMR) is plotted as a function of B-site bismuth doping (x).

The replacement of $Mn^{+3}$ ions by larger $Bi^{+3}$ modulate the trapping of electrons through JT-distortions, therefore alter the DE-mechanism, which may be one of the possible explanations of enhancement of magnetization, $T_C$, MR and AMR in $La_{0.5}Ca_{0.5}Mn_{0.98}Bi_{0.02}O_3$ thin film [3-9]. Our observations in conjunction with the previous results reported by A. Krichene et al. [5-6] concludes that B-site bismuth substitution directly affects the ferromagnetic ground state of LCMBO, thus modulates the DE-mechanism [5-9]. Further, a limited amount of Bi-doping (i.e, $La_{0.5}Ca_{0.5}Mn_{0.98}Bi_{0.02}O_3$) on B-site produces a significant disorder in LCMBO, which improve the electrical and magnetic properties of LCMBO system [5-6].

It will be interesting to investigate the strain effect on $La_{0.5}Ca_{0.5}Mn_{0.98}Bi_{0.02}O_3$ thin film introduced by reducing the thickness of the film [6]. Further, it will be also fascinating to see such change in magnetization and anisotropy can also be observed directly by measuring the magnetization and anisotropy in the presence of applied electric field using insulating or piezoelectric gate [21].

**Conclusions**

In conclusion, we have shown that it is possible to realize the modulation of magnetic (i.e., $T_C$ and magnetization) properties, transport ( i.e., magnetoresistance (MR)) properties and magnetic anisotropy (i.e., anisotropic magnetoresistance (AMR)) of $La_{0.5}Ca_{0.5}Mn_{1-x}Bi_xO_3$ thin films as a function of temperature and magnetic field by replacement of $Mn^{+3}$ ions with larger $Bi^{+3}$ ions. We have observed enhanced AMR for $La_{0.5}Ca_{0.5}Mn_{0.98}Bi_{0.02}O_3$ thin film which may be attributed to the enhancement of magnetic disordering due to the change in unit cell in $La_{0.5}Ca_{0.5}Mn_{0.98}Bi_{0.02}O_3$ thin film. This study is a demonstration of the high-performance materials for high-efficiency spintronics devices based on Manganite (e.g, spintronics based switching devices using anisotropic magnetoresistance (AMR) properties).

**Acknowledgments**

We are grateful for availability of the Institute central facility (SQUID-VSM) in the Department of Physics and IITB Nanofabrication facility (IITBNF) in the Department of Electrical Engineering and Centre of Excellence in Nanoelectronics (CEN), Indian Institute of Technology Bombay.

**References**

[1] Y. Tokura (2000) Colossal magnetoresistive oxides. Gordon and Breach Science Publishers Singapore 2:2

[2] L. B. Freund and S. Suresh (2003) Thin Film Materials. Cambridge University Press 2

[3] H. Gencer, T. Izgi, N. Bayri, M. Pektas, V. S. Kolat, S. Atalay (2016) Structural, Magnetic and Magnetocaloric Properties of $Pr_{0.68}Ca_{0.32-x}Bi_xMnO_3$ (x = 0, 0.1, 0.18, 0.26 and 0.32) Compounds. J Supercond. Nov. Magn. 29:2443

[4] Ram Seshadri and Nicola A. Hill (2001) Visualizing the Role of Bi 6s "Lone Pairs" in the Off-Center Distortion in Ferromagnetic $BiMnO_3$. Chem. Mater. 13:2892

[5] A. Krichene, P.S. Solanki, S. Rayaprol, V. Ganesan, W. Boujelben, D.G. Kuberkar (2015) B-site bismuth doping effect on structural, magnetic and magnetotransport properties of $La_{0.5}Ca_{0.5}Mn_{1-x}Bi_xO_3$. Ceramics International 41:2637

[6] A.Krichene, M.Bourouina, D.Venkateshwarlu, P.S.Solanki, S.Rayaprol, V. Ganesan, W.Boujelben and D.G.Kuberkar (2016) Correlation between electrical and magnetic properties of polycrystalline $La_{0.5}Ca_{0.5}Mn_{0.98}Bi_{0.02}O_3$. J. Magn. Magn. Mater. 408:116

[7] P.E.Schiffer, A.P.Ramirez, W.Bao, S.W.Cheong (1995) Low temperature magnetoresistance and the magnetic phase diagram of La1-xCaxMnO3. Phys.Rev. Lett. 75:3336

[8] H. Sharma, M.R. Lees, G. Balakrishnan, Don McK. Paul, A. Tulapurkar, C.V. Tomy (2017) Electric field controlled magnetization and charge-ordering in Pr0.6Ca0.4MnO3. Mate. Chem. and Phys. 194:142

[9] A.Krichene,W.Boujelben, A.Cheikhrouhou, Structural, electricalandmagnetic study of polycrystalline La0.4Bi0.1Ca0.5MnO3, J.Alloy.Compd. 581(2013) 352.

[10] H. Sharma, A. Tulapurkar and C. V. Tomy (2014) Sign reversal of anisotropic magnetoresistance in $La_{0.7}Ca_{0.3}MnO_3/SrTiO_3$ ultrathin films. Appl. Phys. Lett. 105:222406

[11] J D Fuhr, M Granada, L B Steren and B Alascio (2010) Anisotropic magnetoresistance in manganites: experiment and theory. J. Phys.: Condens. Matter 22:146001

[12] Yali Xie, Huali Yang, Yiwei Liu, Zhihuan Yang, Bin Chen, Zhenghu Zuo, Sadhana Katlakunta, Qingfeng Zhan and Run-Wei Li. (2013) Strain induced tunable anisotropic magnetoresistance in $La_{0.67}Ca_{0.33}MnO_3/BaTiO_3$ heterostructures. J. Appl. Phys. 113:17C716

[13] Y. Z. Chen, J. R. Sun, T. Y. Zhao, J. Wang, Z. H. Wang, B. G. Shen, and N. Pryds. (2009) Crossover of angular dependent magnetoresistance with the metal-insulator transition in colossal magnetoresistive manganite films. Appl. Phys. Lett. 95:132506

[14] Baomin Wang, Lu You, Peng Ren, Xinmao Yin, Yuan Peng, Bin Xia, Lan Wang, Xiaojiang Yu, Sock Mui Poh, Ping Yang, Guoliang Yuan, Lang Chen, Andrivo Rusydi and Junling Wang, Oxygen-driven anisotropic transport in ultra-thin manganite films. (2013) Nat. Commun. 4, 2778.

[15] J. N. Eckstein, I. Bozovic, J. O'Donnell, M. Onellion, and M. S. Rzchowski (1996) Anisotropic magnetoresistance in tetragonal $La_{1-x}Ca_xMnO_\delta$ thin films. Appl. Phys. Lett. 69:1312

[16] M. Egilmez, M. M. Saber, A. I. Mansour, Rongchao Ma, K. H. Chow, and J. Jung, Dramatic strain induced modification of the low field anisotropic magnetoresistance in ultrathin manganite films. (2008) Appl. Phys. Lett. 93:182505

[17] H. Sharma, H. Bana, A. Tulapurkar and C. V. Tomy (2016) Sign reversal of angular dependence of Planar Hall Effect in $La_{0.7}Ca_{0.3}MnO_3/SrTiO_3$ ultrathin film. Mate. Chem. and Phys. 180:5

[18] M. Oumezzine, O. Pena, S. Kallel, N. Kallel, T. Guizouarn, F. Gouttefangeas, M. Oumezzine (2013) Appl. Phys. A. 114:819

[19] S. Jain, H. Sharma, A. K. Shukla, C.V.Tomy, V.R. Palkar, A. Tulapurkar (2014) Optimization of $La_{0.7}Sr_{0.3}MnO_3$ thin film by pulsed laser deposition for spin injection. Physica B. 02:061

[20] H. Sharma, H. Bana, A. Tulapurkar and C. V. Tomy (2016) Planar Nernst and Hall effect in patterned ultrathin film of $La_{0.7}Sr_{0.3}MnO_3$. Mate. Chem. and Phys. 180:390

[21] H. Sharma, A. Tulapurkar and C. V. Tomy (2017) Electric field controlled magnetization and transport properties of La0.7Ca0.3MnO3 ultrathin film. Mate. Chem. and Phys. 186:523